%% file: main.tex
\documentclass[sigconf]{acmart}

\usepackage{bm}
\usepackage{caption}
\usepackage{subcaption}

\usepackage{multirow}
\usepackage{array}
\usepackage{float}

\newcommand{\Zu}[1]{Figure \ref{fig:#1}}

\usepackage{xcolor} %

\newcommand{\erase}[1]{} %

\AtBeginDocument{%
  \providecommand\BibTeX{{%
    \normalfont B\kern-0.5em{\scshape i\kern-0.25em b}\kern-0.8em\TeX}}}

\setcopyright{none}
\renewcommand\footnotetextcopyrightpermission[1]{} %

\begin{document}

\title{Fair Machine Guidance to Enhance Fair Decision Making in Biased People}

\author{Mingzhe Yang}
\affiliation{%
  \institution{The University of Tokyo}
  \city{Tokyo}
  \country{Japan}
}
\email{mingzhe-yang@g.ecc.u-tokyo.ac.jp}

\author{Hiromi Arai}
\affiliation{%
  \institution{RIKEN}
  \city{Tokyo}
  \country{Japan}
}
\email{hiromi.arai@riken.jp}

\author{Naomi Yamashita}
\affiliation{%
  \institution{Kyoto University}
  \city{Kyoto}
  \country{Japan}
}
\email{naomiy@acm.org}

\author{Yukino Baba}
\affiliation{%
  \institution{The University of Tokyo}
  \city{Tokyo}
  \country{Japan}}
\email{yukino-baba@g.ecc.u-tokyo.ac.jp}

\renewcommand{\shortauthors}{Yang et al.}

\begin{abstract}

Teaching unbiased decision-making is crucial for addressing biased decision-making in daily life. Although both raising awareness of personal biases and providing guidance on unbiased decision-making are essential, the latter topics remains under-researched. In this study, we developed and evaluated an AI system aimed at educating individuals on making unbiased decisions using fairness-aware machine learning. In a between-subjects experimental design, 99 participants who were prone to bias performed personal assessment tasks. They were divided into two groups: a) those who received AI guidance for fair decision-making before the task and b) those who received no such guidance but were informed of their biases. The results suggest that although several participants doubted the fairness of the AI system, fair machine guidance prompted them to reassess their views regarding fairness, reflect on their biases, and modify their decision-making criteria. Our findings provide insights into the design of AI systems for guiding fair decision-making in humans.

\end{abstract}


\pagestyle{plain} %
\maketitle

\input{source/body/introduction}

\input{source/body/relatedworks}

\input{source/body/preliminaries}
\input{source/body/experiment}

\input{source/body/results}

\input{source/body/discussion}

\input{source/body/conclusion}

\begin{acks}
This work was supported by JSPS KAKENHI Grant Number 21H03504, JST PRESTO Grant Number JPMJPR19J9 and JST CREST Grant Number JPMJCR21D1, Japan. 
\end{acks}

\bibliographystyle{ACM-Reference-Format}
\input{source/main.bbl}

\clearpage
\appendix
\input{source/body/appendix}

\end{document}

%% file: source/body/introduction.tex
\section{Introduction}
Humans often judge people of particular races and genders unfairly.
In one study, recruiters were sent resumes of people with the same background and skills, differing only by name. Resumes with names commonly associated with white people (e.g., Emily and Greg) were significantly more likely to be called back for job interviews than were those with more prevalent African-American names (e.g., Lakisha and Jamal)~\cite{bertrand2004AreEmily}. Another study revealed that blindfolding judges to hide the gender of participants during orchestral auditions increased the proportion of selected women~\cite{goldin2000American}.

Although educational programs and tools have been developed to address biases when judging people, a recent survey showed that their effects are limited~\cite{Chang2019TheMixed}. In a large-scale survey conducted at a company, the Implicit Association Test (IAT)~\cite{greenwald1998measuring,Dingler2022-ij} was used to assess the employees' unconscious gender biases by measuring the ease with which they could associate words related to gender with those related to career/family. The IAT results were shared with employees to inform them of their biases. A lecture-based training program was implemented to teach employees how to manage gender stereotypes. The post-program results showed a positive effect on their attitudes and an increased awareness of the need to support women. However, the effect on their decision-making was insignificant. Even after administering the course, the selection rate of women as mentors and top employees did not change.

Wilson and Brekke argued that simply being aware of biases and having the motivation to correct them is insufficient to address them. This concept is part of a debiasing framework called the \textit{mental correction process}~\cite{Wilson_Brekke_1994}, which includes four key steps to effectively counter bias: awareness of the bias, motivation to correct it, understanding its direction and magnitude of the bias, and adjusting one's responses accordingly.
According to the large-scale experiment conducted within the company previously described, the lack of improvement in decision-making bias can be understood through this framework. Although the approach of using the IAT and lectures increased awareness and motivation to correct bias, it fell short of offering guidance on how to adjust responses effectively.

Based on these findings, the overarching goal of our research is to develop an AI system that guides humans toward making fair decisions. Although many studies have explored AI-based decision support~\cite{Lai2021-bs,Sah2013-wy,Schultze2015-ue,Onkal2009-ww,He2023-ln,Ma2023-bi,Sivaraman2023-li,Rechkemmer2022-dg,Chiang2023-tm}, our study adopted a different approach. Most existing studies have focused on the impact of AI \textit{assistance} on decision-making. For instance, they have helped users by showing AI-predicted outcomes while assessing the likelihood of someone reoffending~\cite{Wang2021-lk,Chiang2023-tm} or evaluating someone's annual income~\cite{Ma2023-bi}. Users can then use this information for their final judgment. These approaches are similar to the perpetual use of training wheels to assist people in riding bicycles.

In contrast to the existing research, our study employed the mental correction process and aimed to use AI to \textit{guide} people in making better judgments independently without requiring continuous AI assistance. In reference to the aforementioned bicycle analogy, our approach is similar to learning to riding without using training wheels. This human-centered form of decision-making is particularly valuable for personal evaluations, such as face-to-face interactions, wherein AI assistance may not always be feasible. Moreover, studies have shown that humans tend to resist algorithms when making subjective judgments; this phenomenon is called algorithmic aversion~\cite{Schecter2023-ju}. 
Although numerous studies have examined user behavior in AI-assisted decision-making, few have investigated the role of AI in teaching decision-making skills. Only one study examined the effects of AI-based teaching; participants were asked to evaluate the healthiness of foods from their images, and an AI presented evidence for them to consider~\cite{10.1145/3490099.3511138}. Although learning was confirmed, real AI was not used in the experiments, and the study lacked a detailed analysis of the impact of AI on user behavior.

In this study, we developed an AI system to guide individuals in making unbiased decisions. We focused on fair individual evaluations that were closely tied to the users' personal values. We refer to this educational approach as \textit{fair machine guidance}. Although the existing research indicates that humans often exhibit algorithm aversion when AI assistance is involved in such tasks, we investigated whether this aversion also occurs in the context of AI guidance. If not, what processes do users go through to accept AI guidance, and how does this influence their decision-making? To conduct a detailed analysis of how AI guidance impacts human decision-making, we examined the effects of fair machine guidance at each step of the mental correction process. The following research questions were answered:
\begin{itemize}
    \item \textbf{RQ1}: 
    How and to what extent does fair machine guidance affect users' motivation to correct bias?
    \item \textbf{RQ2}: 
    How and to what extent does fair machine guidance affect users' awareness of the magnitude and direction of the bias?
    \item \textbf{RQ3}: 
    How and to what extent does fair machine guidance support users' adjustment to fair decision-making?
\end{itemize}

We used fairness-aware machine learning (ML)~\cite{agarwal18a,Mehrabi2021-lv,Zhang2018-va,G_Harris2020-st} to develop a fair machine guidance system. This approach adjusts ML models to ensure fairness. The process begins by asking the user to evaluate individuals. We then used standard ML techniques, such as logistic regression, to analyze the evaluations and determine the user's decision criteria, that is, the attributes they focused on when making evaluations. Subsequently, we applied fairness-aware ML to the same data to create a fair model and identify the decision criteria when the user made fair assessments. By presenting these results to the users, we aim to help them recognize their own biases and learn to make fair judgments.

We conducted an experiment with 99 participants using our fair machine guidance system. 
The participants were assigned tasks that involved predicting annual income and repayment risk, which served as practical scenarios for human evaluations. In this experiment, we used demographic parity~\cite{Dwork2012-va} as the criterion for fairness, which requires everyone to have an equal chance of receiving a favorable evaluation regardless of sensitive attributes, such as race. We calculated an unfairness score based on this criterion to assess changes in participant biases before and after the educational intervention. To understand the participants' thought processes and how they interacted with the system, we administered questionnaires before and after the intervention. For comparison, we also prepared a baseline method that simply displayed numerical bias metrics, such as ``You evaluated x\% of men as 'high-risk' and y\% of women as `high-risk'.'' We compared this baseline with our fair machine guidance system to explore its impact on various stages of the mental correction process.

The results showed that there was no significant difference in bias reduction between our fair machine guidance method and the baseline method. However, the manner in which participants achieved these improvements differed significantly depending on the method employed. For RQ1, we found that fair machine guidance encouraged users to think more critically about the fairness of their evaluations, prompting them to consider societal expectations of fairness. For RQ2, our system allowed them to visualize their decision-making criteria, which decreased their confidence in the fairness of their judgments. Many participants also realized that their decisions were based on partial information. For RQ3, fair machine guidance led to adjustments in participants' decision-making criteria. Importantly, most participants did not blindly accept the guidance from the system and instead they evaluated the information provided before deciding whether to adapt their approach. Although several participants expressed skepticism, there were cases in which reviewing the AI system’s decision criteria led them to develop new strategies for making fair judgments. 

This paper makes the following contributions:
\begin{itemize}

\item A novel approach, "Fair Machine Guidance", which leverages AI to guide individuals in making unbiased decisions, ensuring that people can continue to make such decisions independently, even in the absence of AI assistance.
\item Empirical results demonstrating the efficacy of fair machine guidance in prompting individuals to reassess their perceptions of fairness, reflect on their own biases, and adjust their decision-making criteria accordingly.
\item Implications for the design of AI systems aimed at reducing implicit biases. In particular, our study underscores the importance of moving beyond merely focusing on the acceptance or rejection of AI suggestions. Instead, it advocates for stimulating critical engagement and self-reflection among their users. 

\end{itemize}

%% file: source/body/relatedworks.tex
\section{Related Work}

\subsection{Bias mitigation}
Several methods have been investigated to mitigate biases in human decision making. The IAT has been used extensively to make people aware of their unconscious biases~\cite{greenwald1998measuring,Dingler2022-ij}. 
Existing studies have used the IAT to measure individual biases against race and gender, indicating the presence of bias in participants~\cite{Chang2020-be, Goedderz2022-wi}. A study found that participants exhibited a change in their motivation to correct their biases once they were made aware of them~\cite{Chang2020-be}.
Another study showed that when the current degree of bias was presented to participants, many participants were defensive and surprised~\cite{Goedderz2022-wi}. After being instructed to modify their judgments, and with their bias subsequently reassessed using the IAT, their surprise decreased. This suggests that human biases are unconscious and that people do not heed them until they are highlighted.

Educational programs have been used to motivate people to modify their biases, and their effectiveness has been studied. A study investigated the impact of online diversity training on attitudes and behaviors toward women after participants shared their pre-measured biases~\cite{Chang2020-be}. The results showed that although their attitudes toward women changed, there was no improvement in actual personnel ratings. Another study provided a video-based course to detect sexism in judgments~\cite{Konig2022-in}. The intervention improved the sensitivity of detecting sexism in judgments. However, the authors concluded that this occurred at the expense of specificity, resulting in no real improvement in sexism detection.
Recently, an attempt was made to develop an educational program using virtual reality (VR). The participants were provided with a VR experience of being homeless, and its impact was investigated~\cite{Asher2018-ok}. The results showed that, in the VR simulation of homelessness, participants were more inclined to act in ways that demonstrated empathy and understanding of the homeless individuals after the experiment.
Another study investigated the behavioral changes that occured when the skin color of an avatar was changed during a VR session~\cite{Peck2013-xf}. A comparison of biases measured using the IAT before and after skin-color replacement in VR indicated a decrease in the participants' racial biases. 
Methods aimed at directly correcting stereotypes in humans, such as measuring unconscious bias using the IAT, participating in educational programs, and using VR, have been found to influence bias awareness and motivation to correct bias. However, these methods have been unsuccessful in influencing response adjustments, that is, behavior toward fair judgments.

These studies asked people to correct stereotypes directly. However, the next approach involved mitigating stereotypes without human awareness. Various user interfaces have been proposed to mitigate unconscious stereotyping~\cite{Leung2020-xl,Ma2022-cr,Peng2019-lg}. An experiment that simulated hiring decisions was also conducted~\cite{Leung2020-xl} and the results revealed bias against African Americans. Adjustments to the interface design, such as providing more information about the candidates or limiting the number of choices, effectively reduced these biases. To investigate racial bias on dating websites, a study created a pseudodating website and asked participants about their preferences based on a person's photo and profile~\cite{Kim2022-qi}. The results showed that even those who explicitly stated that they had no racial bias made biased judgments. To address this bias, the authors changed the order of information presentation and found that the bias was significantly reduced when the profile information was presented first. Although these interfaces resulted in improving user stereotypes, the evaluators’ behavior toward stereotypes did not improve.

\subsection{Human-AI collaboration for fair decision-making}

The use of AI in high-stakes decision-making scenarios, such as loan approvals~\cite{Nakao2022-lu}, the risk of judicial recidivism~\cite{Chiang2023-tm,angwin2022machine}, and child welfare consultations in public administration~\cite{Cheng2022-af,Saxena2021-ke,De-Arteaga2020-in,Grimon_undated-lv}, has been explored. However, reports of unfair AI output have raised concerns~\cite{angwin2022machine,Cheng2022-af}. One reason for the unfair output of AI models is that the data used to train the models contain unfair biases. To mitigate this problem, efforts are being made to incorporate human judgments. For instance, in child welfare screenings, reviewing and selectively using AI outputs rather than directly applying them reduces the racial biases that are observed when using an AI model alone~\cite{Cheng2022-af}. Similarly, a collaboration between humans and machines in the decision-making process has been reported to improve fairness when predicting recidivism risk~\cite{Chiang2023-tm}. Additionally, a framework in which humans interactively assess AI outputs and teach the AI what they consider fair has been proposed~\cite{Nakao2022-lu}.

Therefore, several studies have reported that human intervention in the decision-making process can induce fairness in scenarios where using AI alone may produce unfair results. This could be because human estimation and correction of AI errors lead to fairer judgments~\cite{De-Arteaga2020-in}. While earlier research efforts have largely aimed at minimizing bias present in AI systems, our study shifts the focus towards addressing biases inherent in humans themselves.

\subsection{Human-AI decision making}

Several studies have been conducted on decision-making supported by AI~\cite{Lai2021-bs,Sah2013-wy,Schultze2015-ue,Onkal2009-ww,He2023-ln,Ma2023-bi,Sivaraman2023-li,Rechkemmer2022-dg,Chiang2023-tm,Wang2021-lk}. In these studies, ML models advised evaluators regarding the task that was used to make a decision. They performed various tasks and experimented with even more difficult tasks, such as diagnostic assistance for clinicians~\cite{Bach2023-vo,Panigutti2022-xu,Sivaraman2023-li}. AI-based suggestions have also been investigated using various methods. Those used for presenting information involve enumerating answers or candidate answers~\cite{Tolmeijer2022-fs,Chiang2023-tm}, presenting the reasons for the AI system's decisions~\cite{Sivaraman2023-li,Rechkemmer2022-dg}, and presenting views opposing human predictions~\cite{Bach2023-vo,Rechkemmer2022-dg}. The goal of AI-based decision-making is to improve task results, and it has been demonstrated that the final outcome of a task is better with AI assistance than when performed by humans alone. 

Research on human-AI decision making has focused on assessing the information that is useful for humans when using AI and making them trust it~\cite{poursabzi2021manipulating,hase2020evaluating,nguyen2018comparing,Wang2021-lk,Ma2023-bi,Wang2021-yd,Park2022-sd}. Many studies have shown that presenting the reasoning behind a decision can be useful for building trust in AI. For example, presenting the probability of correctness of the evaluator and AI~\cite{Ma2023-bi} and presenting the accuracy and confidentiality of the model~\cite{Rechkemmer2022-dg} to the evaluator help build this trust. One study investigated whether presenting different types of information had different effects on human understanding~\cite{Wang2021-lk}. They found that if the evaluator was familiar with the task, highlighting important points in the judgment could effectively improve trust in the model. Another study investigated the behavior of humans receiving judgments from AI~\cite{Park2022-sd} and confirmed that when people receive a decision from the AI model, they are not convinced unless the AI model explains the reasoning behind its decision. Furthermore, when they received decisions from the AI model, they felt that they were biased if they were made by the AI model alone. It was also suggested that collaborative decisions made by humans and AI are easier to accept. Other methods for mitigating AI distrust include adjusting the timing of presenting AI information. For example, it has been reported that it is better to present AI information only when a human asks for it~\cite{Gemalmaz2022-iy} or after they have made a decision, prompting them to change their final decision~\cite{Panigutti2022-xu,Chiang2023-tm}.

These AI-based decision support systems do not focus primarily on enhancing evaluators' individual decision-making abilities; instead, they aim to optimize task outcomes by leveraging AI capabilities. In contrast, our study aimed to improve decision-making performance in non-AI situations, with the main goal of improving human decision-making.

\subsection{AI teaching}

Only one study used AI to train humans and focused on the impact of AI teaching on human decision-making~\cite{10.1145/3490099.3511138}. The authors investigated the types of information required to assist humans in making health judgments regarding food and found that providing only the reasoning behind judgments effectively improved decision-making. This improvement can be attributed to the fact that such a focused provision of information prompts participants to reconsider their decision-making processes. This study has several limitations, such as the use of predefined information and a lack of optimal decision-making information tailored to each participant. In addition, this study was not conducted using an actual AI system, and the influence of AI on humans was not analyzed. By contrast, in this study, the presented information was optimized by the AI system for each evaluator, and the impact of AI training on individuals was comprehensively analyzed.

Machine teaching is an ML method that aims to provide optimal education to humans~\cite{liu2017iterative,chen2018Near-Optimal,zhou2018unlearn,Wang2021-yd,Liu2021-ix,Qiu2023DHT,Yeo2019-jh,10.5555/3454287.3454651,Liu2018ICML}. A framework has been proposed that conceptualizes humans as students and an ML model as the teacher. In this framework, the teacher identifies and sequentially provides students with the most effective learning resources~\cite{liu2017iterative,Liu2018ICML}. However, these studies conducted the experiments via simulations without human participants. By contrast, some machine teaching studies have employed humans as students~\cite{chen2018Near-Optimal,zhou2018unlearn,10.5555/3454287.3454651}, and the tasks involved classifying living things such as birds and insects, for which objectively correct answers exist. These studies showed that humans can learn the behaviors of ML models. Our study addresses the subjective task of personal evaluation rather than that for which there is an objectively correct answer, such as biological classifications. This study is also novel in that it investigates how AI-based education affects subjective evaluations.

%% file: source/body/preliminaries.tex
\section{Fair machine guidance}\label{sec:fairness}
\subsection{Overview}
Fair machine guidance aims to educate individuals on making unbiased decisions. This method is based on the mental correction process~\cite{Wilson_Brekke_1994} and focuses on raising awareness of the direction and magnitude of bias and aiding in response adjustment. Fair machine guidance initially prompts users to perform personal assessments to collect their responses. Thereafter, the logistic regression is used to estimate the user's decision criteria, which is referred to as the \textit{student model}. Additionally, a fairness-aware ML method~\cite{agarwal18a} was applied to train the \textit{teacher model}, which is an unbiased modification of the student model. 

Both the student and teacher models were used to select teaching samples that maximized the learning impact. These samples were selected using iterative machine teaching~\cite{liu2017iterative}, which models how a student model is updated when presented with a teaching sample and selects the samples most likely to align the student model with the teacher model. Each teaching sample advises the user by saying ``You evaluated this person as X, but to be fairer, you should evaluate this person as Y.'' 

\begin{figure*}
    \centering
    \includegraphics[width=\textwidth]{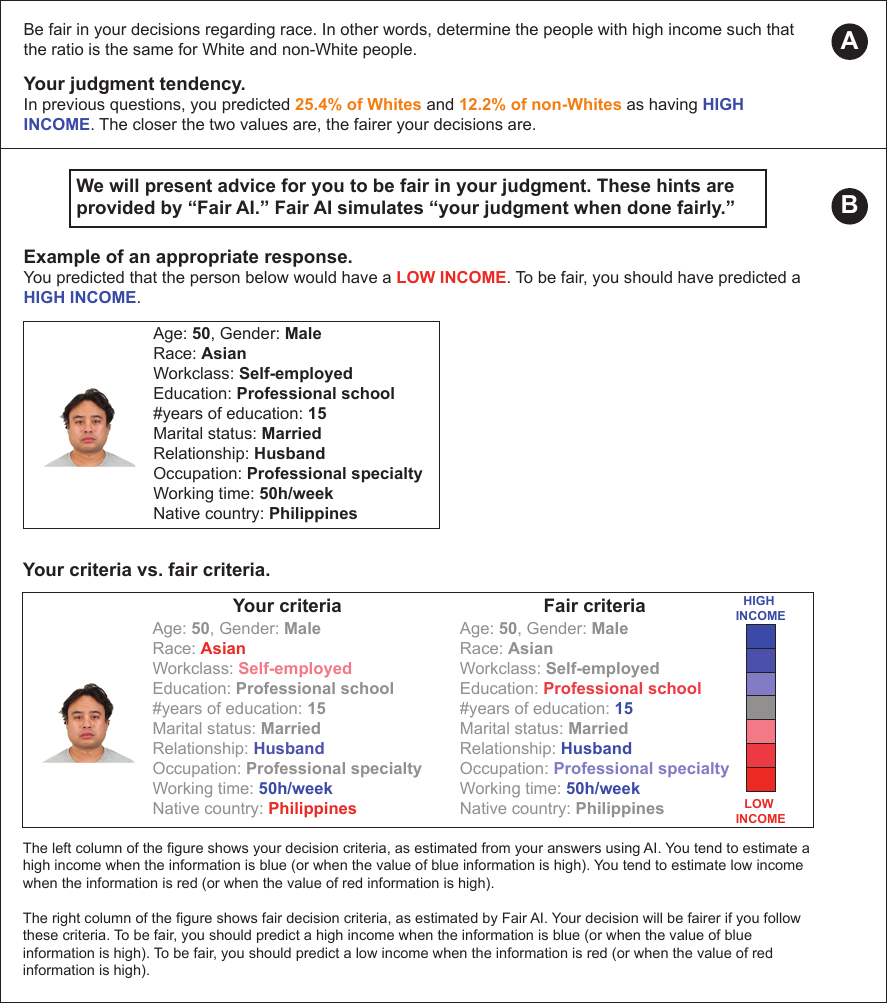}
    \caption{
    Example of fair machine guidance. (A) Current degree of unfairness (i.e., bias feedback); (B) Teaching materials to teach how to make fair decisions. The materials were selected through iterative machine teaching. The interpretations of the student and teacher models were presented visually.
    }
    \label{fig:condition}
    \Description{
    Example of AI based fair machine guidance. (A) Current degree of unfairness (i.e., bias feedback); (B) Teaching materials provided by fair machine guidance to teach how to make fair decisions. The materials were selected through iterative machine teaching. The interpretations of the student and teacher models were presented visually.
    }
\end{figure*}

Fair machine guidance not only provides teaching samples but also offers interpretations of the student and teacher models. The interpretation of the student model is presented visually, allowing users to review their decision-making criteria and understand their biases. The interpretation of the teacher model has also been shown to guide users in adjusting their decision criteria to achieve fairer outcomes. An example interface for fair machine guidance is displayed in Figure~\ref{fig:condition}. In the visualization, the top five attributes with the highest absolute weight values are highlighted.

\subsection{Fairness metrics}\label{subsec:fairness}
A fairness metric is necessary for applying fairness-aware ML. We selected \textit{demographic parity}~\cite{Dwork2012-va}, which is a commonly used fairness metric that aims to ensure that the likelihood of a certain outcome is not influenced by membership in a sensitive group, such as race or gender. For instance, demographic parity is achieved when the probability of hiring a man is the same as that of hiring a woman. In our experiments, we focused on apartment rentals and credit risk assessment. Various national and regional laws legally mandate demographic parity as a fairness metric in these areas\footnote{\url{https://www.justice.gov/crt/fair-housing-act-1}}\footnote{\url{https://www.justice.gov/crt/equal-credit-opportunity-act-3}}\footnote{\url{https://ec.europa.eu/commission/presscorner/detail/en/MEMO_07_257}}. This led us to adopt demographic parity in our experiments. Indeed, other studies using the same datasets as our experiments have also employed demographic parity as a fairness metric~\cite{Jiang_Han_Fan_Yang_Mostafavi_Hu_2022, Liao_Naghizadeh_2023, calders2010three}.

Demographic parity is represented by the equation: $\mathrm{Pr}[\hat{y} \mid z=1] = \mathrm{Pr}[\hat{y} \mid z=0]$, where a sensitive attribute (such as race or gender) is denoted by $z \in \{0, 1\}$. 
We focused on cases wherein $z$ is binary: $z=1$ indicated that the individual belongs to a privileged group, whereas $z=0$ indicated that they belong to an unprivileged group. 
The decision, such as whether to hire or extend credit to someone, is denoted by $\hat{y} \in \{0, 1\}$.
We considered that $\hat{y}=1$ represents a favorable outcome.

\subsection{Fairness-aware ML}
We applied the reduction approach~\cite{agarwal18a} as the fairness-aware ML method. This approach satisfies fairness constraints (such as demographic parity) while reducing classification loss by adding constraints based on fairness criteria as penalty terms to the loss function.
This approach can easily overcome the trade-off between accuracy and fairness. 
We used the PyTorch implementation called FairTorch\footnote{\url{https://github.com/wbawakate/fairtorch}}.

\subsection{Iterative machine teaching}
We applied the iterative machine teaching method~\cite{liu2017iterative} to select efficient teaching materials. 
Suppose that both the teacher and student make decisions according to prediction model $f_{\bm{w}}(\bm{x})$, which is parametrized by a certain parameter $\bm{w}$. 
Let the teacher model be $\bm{w}_*$ and the student model after the $t$-th training be $\bm{w}_t$. 
Assume that $\bm{w}_t$ is updated according to stochastic gradient descent when a certain teaching material, $(\bm{x}, y)$, is given, that is, $ \bm{w}_{t+1}=\bm{w}_t-\eta \frac{\partial \mathcal{L}\left( f_{\bm{w}}(\bm{x}), y\right)}{\partial \bm{w}}$, where $\eta$ is the learning rate and $\mathcal{L}(\cdot, \cdot)$ is the loss function.
In iterative machine teaching, a material $(\bm{x}, y)$ is selected from a set of teaching materials $\Phi$ such that $\bm{w}_{t+1}$ is closest to $\bm{w}_*$. 
In other words, the teaching material $(\bm{x}, y)$ that minimizes $\left\| \bm{w}_{t+1}-\bm{w}_*\right \|_{2} ^{2} = \left\| \bm{w}_{t}-\eta \frac{\partial \mathcal{L}\left( f_{\bm{w}}(\bm{x}), y\right)}{\partial \bm{w}} - \bm{w}_* \right\|_{2}^{2}$ is selected.

%% file: source/body/experiment.tex
\section{Experiments}

\subsection{Experimental design}
We employed a between-subjects study design with two conditions. This study was reviewed and approved by the Ethical Review Committee for Experimental Research involving Human Subjects at the University of Tokyo. 

\subsubsection{Tasks}\label{'experiments_tasks'}
\begin{figure*}
     \centering
     \begin{subfigure}[t]{0.49\textwidth}
         \centering
         \includegraphics[width=.8\textwidth]{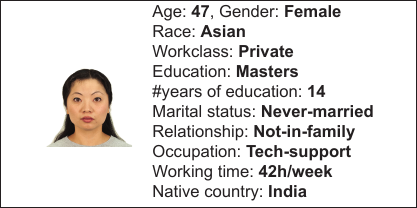}
         \caption{Income}
         \label{fig:problems-income}
     \end{subfigure}
     \hfill
     \begin{subfigure}[t]{0.49\textwidth}
         \centering
         \includegraphics[width=.8\textwidth]{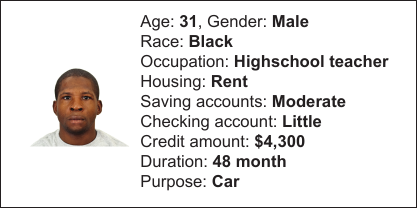}
         \caption{Credit}
         \label{fig:problems-credit}
     \end{subfigure}
        \caption{Examples of synthesized profiles}
        \label{fig:problems}
    \Description{
    In each profile, a photograph of the person's face is displayed on the left, while personal information is displayed on the right.
    }
\end{figure*}

We prepared two personal assessment tasks: Income and Credit. In each task, synthesized person profiles were presented, and the participants were asked to assess the individuals. Examples of these profiles are shown in Figure~\ref{fig:problems}.

\begin{itemize}
    \item \textbf{Income}: 
    In this task, participants were asked to rate their person's annual income as high or low. The personal profiles were generated using the Adult dataset~\cite{misc_adult_2}, which was created using the U.S. census data. In addition to age, race, and gender, the data also included information regarding education, occupation, and working hours. The participants were informed that the top $20\%$ of individuals had the highest annual incomes. This threshold was set based on the ground-truth label ratio.
    \item \textbf{Credit}: 
    In this task, participants were asked to classify the person as high or low credit risk. The German credit dataset was used for generating the profiles~\cite{misc_statlog_german_credit_data_144}. The data included borrower profile (e.g., gender, age, and occupation), credit amount, duration, and purpose. The participants were informed that $30\%$ of these individuals had a high credit risk. This threshold was set based on the ground-truth label ratio. We modified the occupational and housing attributes of the original data to incorporate the gender inequality. We expected fair machine guidance to teach humans to ignore these attributes.
\end{itemize}

In this experiment, $300$ profiles were used for each task. 
For the Income task, the ground-truth labels for high and low income were used for annual incomes of above and below $50{,}000$ USD, respectively. For the Credit task, the results of the actual risk assessment were provided as ground-truth labels. 
We sampled the profiles to ensure that the distribution of the ground-truth labels was equal for both privileged and unprivileged groups. 
Additionally, the ground-truth labels were only accessed during data sampling.

Race (white or non-white) and gender (male or female) were used as sensitive attributes for the Income and Credit tasks, respectively. In the Income task, $37.3\%$ of the profiles were of white and $62.7\%$ were non-white people. In the Credit task, $47.3\%$ of the profiles were males and $52.7\%$ were females. In the Income task, the participants were instructed to assume the role of a landlord who wanted to lease a house to someone with a high income and thereby scrutinize the prospective tenant's profile to predict and select a person with a high income. Moreover, the participants were instructed to be fair regarding race; i.e., they were instructed to select the same percentage of people with high incomes from both groups. In the Credit task, the participants assumed the role of credit examiners. The participants were instructed to scrutinize the profiles of the loan applicants, and anticipate and select those who were less likely to repay the loan (i.e., ``high risk''). They were instructed to be fair in terms of gender.
To emulate a face-to-face assessment, a photograph of the person's face was presented along with their profile.
We used photos of Asian, Black, and White female and male models from the Chicago Face Database~\cite{ma2015chicago}. %

\begin{table*}[t]
\caption{Sensitive attributes and instructions for in-person assessment tasks}\label{tb:tasks}
\small
\begin{tabular}{|l|p{2.2cm}|p{10cm}|}
\hline
Task & Sensitive attribute & Instructions \\
\hline
\hline
Income & Race (White or Non-White) & 
Assume that you are the landlord of an apartment building. You have received a rent request from someone who lives in the United States. You want to lease the apartment to someone with a high income. You have been instructed to scrutinize the prospective tenant's profile and predict their annual income. Please predict a person with a ``high income'' based on the information provided. A person among the top 20 percentile of earners is that with ``high income.'' However, you have been instructed to judge each race fairly. In other words, please ensure that the proportion of White and non-White with high annual incomes is the same. Please answer such that both the rate of correct answers and level of fairness are high. 
\\
\hline
Credit & Gender (Female or Male) & 
Assume that you are working in a financial institution and are conducting the credit screening a loan applicant. You have been instructed to predict the probability of loan repayment by the applicant. Please predict a person who is ``high risk'' (i.e., low probability of repayment) based on the information provided. Of the total population, $30\%$ are ``high risk.'' However, you have been instructed to judge each gender fairly. Therefore, please ensure that the proportion of ``high risk'' people is the same across both genders. Please answer such that both the rate of correct answers and level of fairness are high. 
\\
\hline
\end{tabular}
\end{table*}

\subsection{Participants}
The participants were recruited through the crowdsourcing platform Lancers\footnote{\url{https://www.lancers.jp/}}. A total of $459$ participants were asked to complete a pre-test involving $100$ personal assessments. To target only those participants making biased judgments, their unfairness score (defined in Section~\ref{sec:unfairness}) was measured for the pre-test, and only those who scored $0.03$ or higher progressed to the next phase. The threshold was determined using the distribution of unfairness obtained through preliminary experiments conducted with different participants. In the preliminary experiment, we asked the participants to perform the same person assessment tasks, but did not instruct them to be fair. According to the distribution of unfairness scores obtained in the preliminary experiment, the top, middle, and bottom 1/3 of the participants were divided into biased, normal, and reverse-biased groups, respectively. Individuals in the biased group had unfairness scores of approximately $0.03$ or higher.

Owing to the performance limitations of fairness-aware ML, there were cases in which the obtained fair model was less fair than the judgments of some participants. These participants were excluded at the pre-test stage. Although the fairness model was used only in the fair machine guidance condition, the same screening was applied in the bias feedback condition to ensure consistency among the participants. A total of $99$ participants proceeded to complete the main experiment.

According to the post-experiment survey, of the $99$ participants, $50$ were female, $48$ were male, and $1$ was non-binary. Nine participants were aged $20$--$29$ years, $26$ aged $30$--$39$ years, $37$ were aged $40$--$49$ years, $22$ were aged $50$--$59$ years, and $4$ were aged $>60$ years. Only one person preferred not to indicate their age. Finally, all the participants were of Asian origin.

\subsection{Procedure}\label{sec:procedure}
The experiment comprised of four phases: pre-test (as previously described), treatment, mini-test, and post-test. After completing the pre-test, the participants completed a questionnaire before moving on to the treatment phase. During this phase, different treatments were applied to each condition, as described in Section~\ref{sec:conditions}. Thereafter, the participants assessed $20$ people on a ``mini-test''. This cycle of treatment and mini-test was repeated five times. Finally, a post-test was conducted, wherein the participants assessed $100$ profiles. The same profiles were used for the pre-test, mini-tests, and post-test evaluation in each task. After completing the post-test, the participants completed an additional questionnaire. This procedure is illustrated in \Zu{workflow}. 

Participants were paid $2.5$ USD to complete the pre-test and an additional $7.5$ USD to complete the remainder of the experiment. The average time for completing all procedures was $47$ min. To encourage serious engagement, we informed them that they would receive extra compensation $7.5$ USD if their assessments met or exceeded certain accuracy and fairness criteria. We offered an additional $7.5$ USD to participants whose accuracy ranked in the top $10\%$.

\begin{figure}[tb]
    \centering
    \includegraphics[width=.75\linewidth]{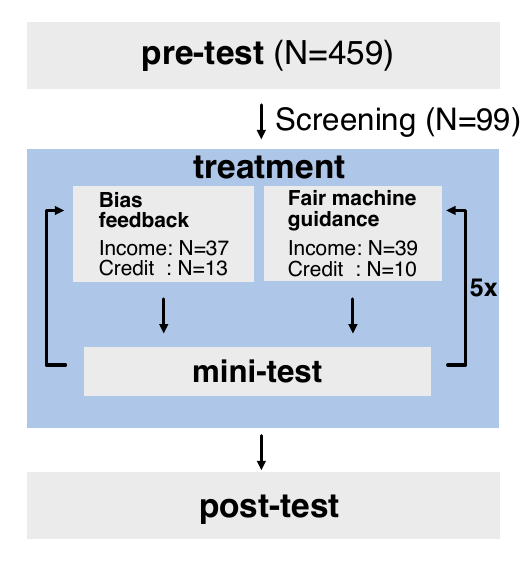}
    \caption{Overview of task processing: Initially, participants completed a pre-test and were screened for the next step based on their unfairness scores. Subsequently, they were randomly assigned to each condition (bias feedback or fair machine guidance) and moved to the treatment phase. This phase included the treatment and mini-test, and this cycle was repeated five times. Finally, they underwent a post-test.}
    \Description{The flowchart provides an overview of task processing in the study. Participants initially underwent a pre-test, followed by a screening based on unfairness scores. They were then randomly assigned to conditions (bias feedback or fair machine guidance) and entered the treatment phase, which included repeated five times cycles of treatment and mini-tests. The process concluded with a post-test.}
    \label{fig:workflow}
\end{figure}

\subsection{Conditions}\label{sec:conditions}
We prepared a simple baseline, referred to as \textit{bias feedback}, which numerically displayed each participant's bias. This baseline was designed to aid in the first step of the mental correction process, which is raising bias awareness; however, it did not offer any guidance on making fairer decisions. By comparing the results with those of the baseline, we examined the effect of fair machine guidance on the subsequent steps of the mental correction process.
\begin{itemize}
    \item \textbf{Bias feedback}: The current level of each participant's unfairness was displayed and was calculated based on their personal assessments in the pre- and mini-tests. The participants were presented with their frequency of selecting individuals from privileged and unprivileged groups. For example, in the Income task, participants were told ``You predicted $25.4\%$ of Whites and $12.2\%$ of non-Whites as having high income.'' Additionally, they were informed that the closer these values were, the fairer their decisions would be. 
    \item \textbf{Fair machine guidance}: In addition to informing participants about their current level of unfairness, each participant was provided teaching materials, as outlined in Section~\ref{sec:fairness}. Both the student and teacher models were personalized and trained by using each participant's individual responses. After the mini-test, the models were updated based on the participants' answers. Teaching samples were selected from the questions that each participant had already answered, and five samples were displayed simultaneously. The participants were also required to complete a check test to ensure that they understood how to interpret the visualizations provided by the model. Only those who passed this test were included in the main experiments.
\end{itemize}

Of the $99$ participants, $37$ were assigned to [Income, Bias feedback], $39$ to [Income, Fair machine guidance], $13$ to [Credit, Bias feedback], and $10$ to [Credit, Fair machine guidance].

\subsection{Data analyses}
\subsubsection{Quantitative analysis}\label{sec:unfairness}
As discussed in Section~\ref{sec:fairness}, we used demographic parity as the fairness metric. We introduced an \textit{unfairness score}, defined as $\mathrm{Pr}[\hat{y}=1 \mid z=1] - \mathrm{Pr}[\hat{y}=1 \mid z=0]$, to gauge unfairness according to demographic parity.
In this context, a positive unfairness score suggests that an evaluator is more likely to make favorable decisions toward individuals in the privileged group ($z=1$) than toward those in the unprivileged group ($z=0$).
We used the \textit{key attribute change rate} between the pre- and post-tests to measure the degree of modification in the decision criteria. After both tests, we asked the participants to select at most five attributes (e.g., age and occupation) that were important for the evaluation. We calculated the similarity between the sets of selected attributes using the Jaccard coefficient and used the $1 - $ Jaccard coefficient to indicate the key attribute change rate. The sets of attributes selected in the pre- and post-tests were denoted as $A_{\mathrm{pre}}$ and $A_{\mathrm{post}}$, respectively, and the key attribute change rate was defined as $1 - \frac{\lvert A_{\mathrm{pre}} \cap A_{\mathrm{post}} \rvert }{\lvert A_{\mathrm{pre}} \cup A_{\mathrm{post}} \rvert}$.

To answer the research questions, we investigated the following subjective measures using a five-point Likert scale (1: Disagree, 5: Agree): 
\begin{itemize}
    \item Reconsideration of fairness (Q10): We measured whether the advice given by the machine influenced participants' perception of fairness.
    \item Self-reported fairness (Q4): We assessed the participants' confidence in the fairness of their answers.
    \item Willingness to follow machine advice (Q5): we measured the extent to which participants were receptive to machine advice.
\end{itemize}

\subsubsection{Qualitative analysis}
In addition to the quantitative analysis, we also asked participants open-ended questions such as ``Q7. Do you think that the `Fair AI' was fair?'' (RQ1); ``Q4. Do you think your decisions were fair?'' (RQ2); and ``Q12. Did your judgment criteria change after receiving the advice?'' (RQ3). The entire questionnaire is provided in the Appendix, in which Q6, Q7 and Q10 are associated with RQ1; Q4 and Q11 are associated with RQ2; and Q5, Q8, Q9 and Q12 are associated with RQ3.

To understand how the system influenced participants' recognition of own biases and their willingness to address these biases, we employed thematic analysis on their survey responses. An author initially developed codes inductively to compile a codebook, which featured codes such as "AI fairness," "self-reported fairness," insights on decision-making criteria, "reliability and acceptance of machine advice," and "usage of machine advice." Following this codebook, two authors coded the responses. Any disagreements in coding were resolved through collaborative discussion among the authors.

%% file: source/body/results.tex
\section{Findings}
\subsection{Effects of improving unfairness}\label{subsec:unfairness}

In this study, two types of tasks were conducted to confirm that the effects of the experiments were not task dependent. Because no significant differences were identified in the results of either task, the results of the two experiments are reported together in subsequent sections. 

\Zu{task_unfairness} shows the scatter plots of the pre- and post-test unfairness score. Both fair machine guidance and bias feedback mitigated unfairness among many participants. There were no significant differences in improvement between the methods. However, a comparison of the survey responses of the participants confirmed different behaviors for leading to changes in their decisions between the methods. In response to RQs $1$--$3$, we performed both quantitative and qualitative analyses of the differences in behavior, and examined the impact of fair machine guidance.

\begin{figure}[tb]
    \centering
    \includegraphics[width=\linewidth]{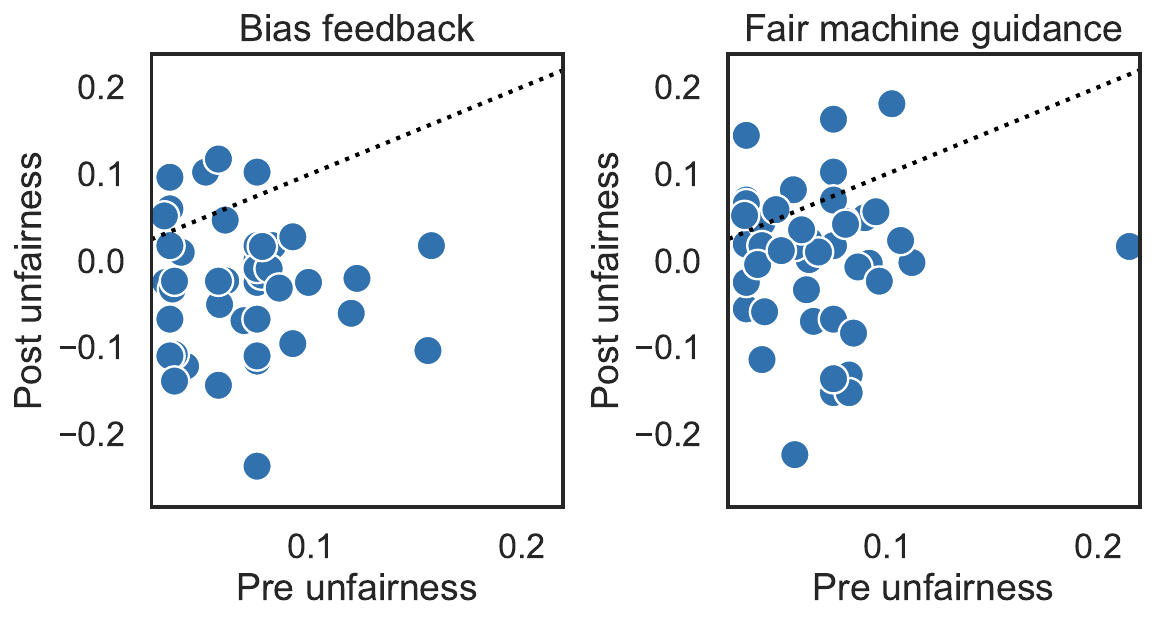}
    \caption{Scatter plots of unfairness for each participant in the pre- and post-tests. The dotted line represents equal levels of unfairness in both the pre- and post-tests; points below this line indicate an improvement in fairness in the post-test.}
    \label{fig:task_unfairness}
    \Description{
    The graph is a scatter plot comparing the bias feedback participants on the left and the fair machine guidance participants on the right. It's entitled "Unfairness of each participant in pre-test and post-test". The horizontal X-axis represents each participant's unfairness in the pre-test. The vertical Y-axis represents each participant's unfairness in the post-test. A dotted line indicates where the unfairness in the pre-test and the post-test are equal. Points below this line represent an improvement in fairness in the post-test. It's evident that many participants in both bias feedback and fair machine guidance groups are below this line, indicating an improvement in fairness.
    }
\end{figure}

\subsection{RQ1: Effects on motivation to mitigate biases}
To answer RQ1, we quantitatively analyzed how fair machine guidance affected the participants' motivation to modify their biases. \Zu{Q10_hist} shows the distribution of responses to Q10 (``Did this task cause you to reconsider the fairness of your own decision and that required by society?'') on a five-point Likert scale. The Mann--Whitney U test revealed a statistically significant difference between fair machine guidance and bias feedback responses, with $p<0.05$. In other words, we confirmed that fair machine guidance was more effective at prompting the reconsideration of fairness and motivating bias correction among the participants.

\begin{figure}[tb]
    \centering
    \includegraphics[width=.8\linewidth]{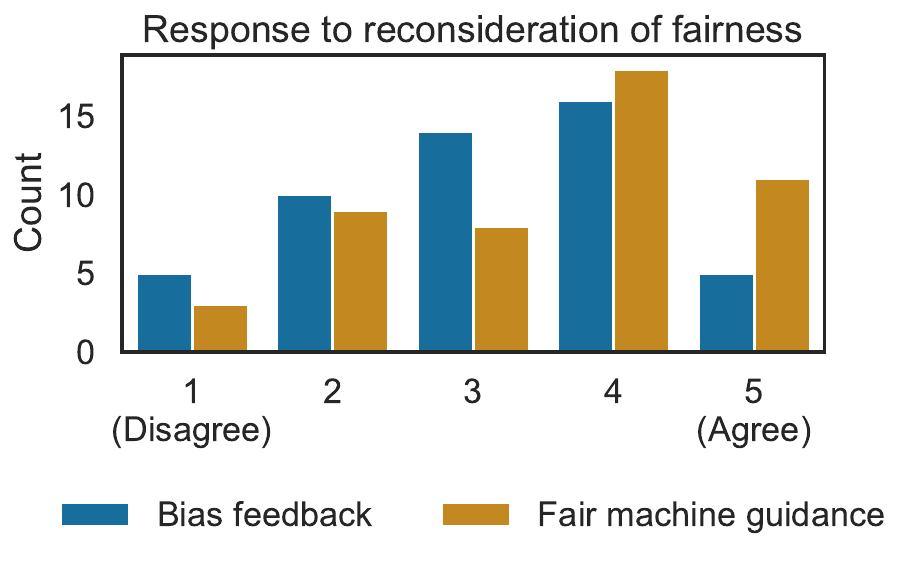}
    \caption{Distribution of responses to Q10: ``Did these tasks cause you to reconsider the fairness of your own decision and that required by society?''}
    \label{fig:Q10_hist}
    \Description{
    The graph is a double bar graph representing the distribution of responses to Q10: "Did these tasks cause you to reconsider the fairness of your own decision and that required by society? The responses were ranked on a 5-point scale, with 1 being 'disagree' and 5 being 'agree'. For both bias feedback and fair machine guidance, a significant number of participants agreed that the task made them reconsider fairness. The distribution of responses is as follows:
    1: Bias feedback 5; Fair machine guidance 3. 
    2: Bias feedback 10; Fair machine guidance 9. 
    3: Bias feedback 14; Fair machine guidance 8. 
    4: Bias feedback 16; Fair machine guidance 18. 
    5: Bias feedback 5; Fair machine guidance 11. 
    It's clear from the graph that many participants in both groups tended to reconsider fairness, especially with higher ratings of 4 and 5.
    }
\end{figure}

Next, we used open-ended responses to qualitatively analyze the effect on motivation for bias correction. First, in the fair machine guidance group, 8 of 49 participants were convinced that the fairness presented by the AI was accurate and that they realized their own bias through this guidance; these participants provided the following responses:
\begin{quote}
    \begin{description}
        \item[P3:] I wanted to make a fair decision based on the AI system's suggestions (hence, I followed them). I did not intend to apportion income by gender; however, I was reminded that this was the basis of my thinking and felt that I had to revise it. [Income, Fair machine guidance]
        \item[P8:] I was unknowingly making biased decisions; therefore, I decided to follow the AI system's suggestions. I thought I was making fair decisions, but realized that I was unknowingly making biased ones, which gave me the opportunity to rethink them. [Income, Fair machine guidance]
    \end{description}
\end{quote}

In contrast, some participants questioned the fairness offered of the AI system. Eleven out of forty-nine participants interpreted the fairness taught by the AI as that required by society, where there were discrepancies between the fairness metric and their own ideas about fairness; 3 of them concluded that it was important to consider things from a wide range of perspectives:
\begin{quote}
    \begin{description}
        \item[P21:] There were instances where the AI system did not make a fair decision, specifically when deciding on annual income according to the country of origin. I felt that it is important to evaluate people from a range of perspectives, rather than based on a single piece of information, such as gender, age, or race. [Income, Fair machine guidance]
        \item[P24:] Unable to state that the AI was impartial because there were both convincing and unconvincing parts the information presented by the AI. I also did not feel that it was fair to predict that an applicant would have high income if they were from Taiwan or low income if they were from the U.S. I felt that even when I intended to be fair, there was an unintentional blurring of standards and I was predisposed to offer preferential treatment in my efforts to be fair. I found it very difficult to find a balance, and it made me think what ``fairness'' is in the first place. [Income, Fair machine guidance]
    \end{description}
\end{quote}

Meanwhile, 9 of the 49 participants lost their understanding of fairness because of fair machine guidance. Nonetheless, some also gained insights into their own preconceptions.
\begin{quote}
    \begin{description}
        \item[P10:] I am not sure that the AI system is truly impartial. Additionally, I am unsure of how to define fairness. I did not intend to discriminate based on race but felt that I was making assumptions regarding education and occupation. [Income, Fair machine guidance]
        \item[P45:] I wonder whether the term ``fairness'' could be so vague and unclear, leaving me feeling unsettled or unsatisfied. [Income, Fair machine guidance]
    \end{description}
\end{quote}

Thus, fair machine guidance provides an opportunity to reconsider the fairness standards required by society. A few participants expressed concerns that the AI system would imprint its impartiality on humans ---``This AI system created by highly educated people appears to have a biased impartiality'' (P70, [Credit, Fair machine guidance]). However, no such responses were observed in the bias feedback group. Some participants stated that the fair machine guidance did not prompt them to reconsider their decision, stating ``Since the decision criteria of AI were limited, I do not think this alone would trigger a reconsideration of the fairness required in society'' (P87, [Income, Fair machine guidance]).

Meanwhile, participants in the bias feedback group exhibited strong opposition to the fairness metric employed in the experiments. In particular, there was significant opposition (14 out of 50) to optimizing demographic parity and resistance to quantifying and adjusting for fairness. However, the antipathy effect was not observed in the group with fair machine guidance.
\begin{quote}
    \begin{description}
        \item[P80:] I believe that a person should receive a favorable decision if they can afford to pay regardless of their race, and that the percentage should not be adjusted based on race. [Income, Bias feedback]
        \item[P31:] It was troublesome to adjust the numbers while considering race, and the fact that the numbers were based on race rather than ability made me wonder, ``What is fairness in the first place?'' [Income, Bias feedback]
    \end{description}
\end{quote}

In addition to the backlash against the impartiality of the bias feedback, 5 of 50 participants were not prompted to reconsider their impartiality because they were confident that their decisions were impartial:
\begin{quote}
    \begin{description}
        \item[P94:] I am fair because I made a comprehensive decision. The AI guidance appeared to provide superfluous information for decision making. [Credit, Bias feedback]
        \item[P54:] I think I would have made the same choice without being presented with ``fairness in choosing race,'' which did not prompt me to reconsider. [Credit, Fair machine guidance]
    \end{description}
\end{quote}
Many participants in the fair machine guidance group stated that they were unsure about fairness, but only a few in the bias feedback group felt this.

Thus, in many cases, bias feedback did not result in motivation owing to confidence in one's own fairness or a backlash because one's own idea of fairness differed from that of the AI system. A comparison between the bias feedback and fair machine guidance groups confirmed that fair machine guidance, which encourages people to review their own decision-making criteria and adjust them according to the fairness metric, is effective at motivating them to address bias by reconsidering the fairness standard required by society and making them aware of their own biases.

\subsection{RQ2: Effects on awareness of direction and magnitude of biases}
To answer RQ2, we quantitatively examined how the participants felt about their decisions after receiving advice from the AI system. The self-assessment results for decision fairness in the post-test are shown in \Zu{Q4_hist}. The Mann-Whitney U test showed a $p<0.05$, which indicated a statistically significant difference between fair machine guidance and bias feedback. Fair machine guidance decreased confidence as participants reviewed their decisions, whereas bias feedback resulted in a trend wherein confidence did not change after receiving guidance.

\begin{figure}[tb]
    \centering
    \includegraphics[width=0.8\linewidth]{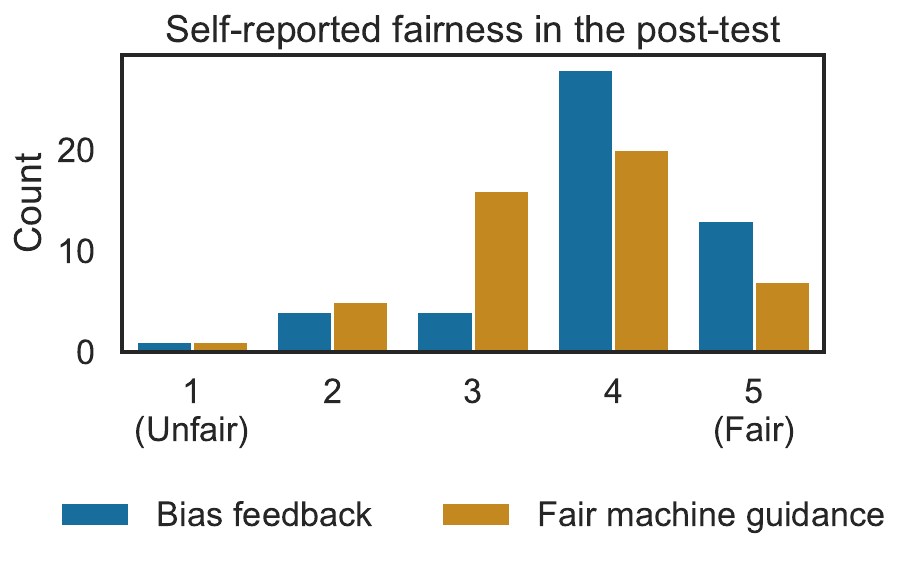}
    \caption{Distribution of responses to Q4: ``Do you think your decisions were fair in the last 10 questions you answered?''}
    \label{fig:Q4_hist}
    \Description{
    The graph is a double bar graph representing the distribution of responses to Q4: "Do you think your decisions were fair in the last 10 questions you answered?" Responses were given on a 5-point scale, with 1 being 'unfair' and 5 being 'fair'. The data are as follows: 
    1: Bias feedback 1; Fair machine guidance 7. 
    2: Bias feedback 4; Fair machine guidance 5. 
    3: Bias feedback 4; Fair machine guidance 16. 
    4: Bias feedback 28; Fair machine guidance 20. 
    5: Bias feedback 13; Fair machine guidance 7. 
    The graph shows that participants in the fair machine guidance group were generally less confident in the fairness of their decisions than those in the bias feedback group, particularly in the middle ratings.
    }
\end{figure}

Next, based on the open-ended comments, we qualitatively analyzed what the participants noticed regarding the magnitude and direction of the bias. Many people (19 of 49) specifically noticed their own biases in the decision-making criteria through fair machine guidance, providing the following responses:
\begin{quote}
    \begin{description}
        \item[P13:] At first, I did not understand why I was wrong, but after being corrected several times, I realized that I was unconsciously making decisions based on factors such as my educational background. As I progressed through the questions, I realized that I was making decisions based on educational background and job title. [Income, Fair machine guidance]
        \item[P74:] This was an opportunity to revise my preconceived notions, especially regarding the country of origin and gender. [Income, Fair machine guidance]
        \item[P15:] I noticed a tendency to judge a person's ability to pay based on their occupation. [Credit, Fair machine guidance]
    \end{description}
\end{quote}

In the bias feedback, only three participants noticed bias in specific criteria:
\begin{quote}
    \begin{description}
        \item[P12:] I have never judged a person's wealth or credibility based on the color of their skin, but in this case, I realized that I was only looking at their educational background. [Income, Bias feedback]
        \item[P32:] I realized that fairness does not depend solely on race. In retrospect, I realized that I held a bias against occupation and education. [Income, Bias feedback]
    \end{description}
\end{quote}

Fair machine guidance also helped participants realize that they had made decisions based on limited information. Those who achieved this realization were not included in the bias feedback group.
\begin{quote}
    \begin{description}
        \item[P21:] I feel that, when evaluating a person, it is important to evaluate them from a range of perspectives rather than judge them based on a single piece of information such as gender, age, or race. [Income, Fair machine guidance]
        \item[P90:] I thought it would be fairer to make decisions based on overall factors, without considering race, etc. [Income, Fair machine guidance]
    \end{description}
\end{quote}
P90 questioned the fairness of the AI system. Despite this, they recognized the importance of viewing diverse information.

Thus, fair machine guidance was found to reduce confidence in the correctness of one's own evaluation by reviewing one's own decision-making criteria, making them aware of their bias against specific decision-making criteria, and helping them realize that they were making decisions based only on partial information.

\subsection{RQ3: Effects of adjusting decision criteria}
\begin{figure}[tb]
    \centering
    \includegraphics[width=0.8\linewidth]{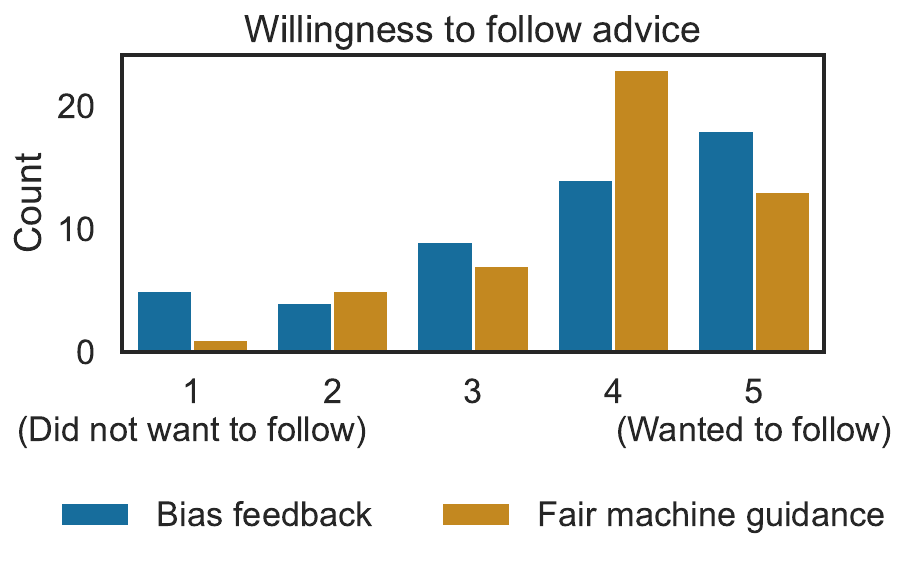}
    \caption{Distribution of responses to Q5: ``We presented the machine-generated advice to help you be fair in your judgment. Did you decide to follow the advice?''}
    \label{fig:Q5_hist}
    \Description{
    The graph is a double bar graph presenting the distribution of responses to Q5: "We presented the machine-generated advice to help you be fair in your judgment. Did you decide to follow the advice?" Participants rated their responses on a 5-point scale, with 1 meaning 'did not want to follow' and 5 meaning 'wanted to follow'. The distribution was as follows:
    1: Bias feedback 5; Fair machine guidance 1.
    2: Bias feedback 4; Fair machine guidance 5.
    3: Bias feedback 9; Fair machine guidance 7.
    4: Bias feedback 14; Fair machine guidance 13.
    5: Bias feedback 18; Fair machine guidance 23.
    The graph showed that a significant number of participants in both the bias feedback and fair machine guidance groups were tending to follow the machine's advice.
    }
\end{figure}

\begin{figure}[tb]
    \centering
    \includegraphics[width=0.8\linewidth]{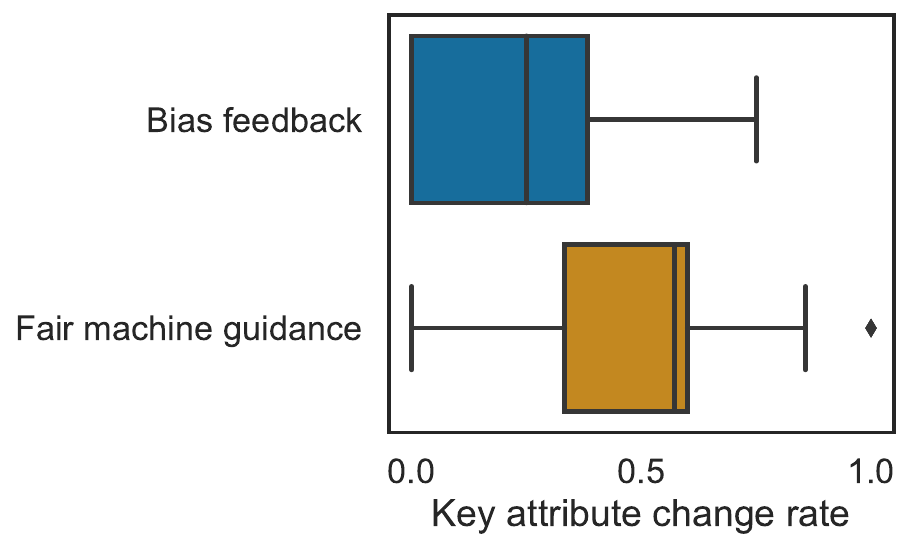}
    \caption{Degree of change in attributes of focus between pre- and post-tests}
    \label{fig:change_feat_rate}
    \Description{
    The box plots represent the degree of change in attention attributes between pre-test and post-test for the two experimental conditions: bias feedback and fair machine guidance. For the fair machine guidance condition, the average rate of change is 0.47, indicating a significant shift in attention attributes. On the other hand, the bias feedback condition has an average rate of change of 0.23, suggesting a tendency for focus attributes to change less.
    }
\end{figure}

To answer RQ3, we examined whether participants accepted the AI guidance and changed their decisions. \Zu{Q5_hist} shows the distribution of the responses to Q5 ``We presented the machine-generated advice to help you be fair in your judgment. Did you decide to follow the advice?''. Previous studies on decision support by AI assistants have shown that participants who strongly relate to personal values, such as personal assessment, are prone to algorithm aversion. However, in the guidance approaches of this study, many participants accepted the advice of the AI system. There was no statistically significant difference in the willingness to follow the advice between the fair machine guidance and bias feedback groups.

Next, we compared the rate of change in the attributes focused on in the pre- and post-tests to determine whether the participants' decision-making criteria changed in response to the guidance; the results are shown in \Zu{change_feat_rate}. The Mann--Whitney U test yielded $p<0.05$, confirming that fair machine guidance was statistically significantly more effective than bias feedback in promoting changes in decision-making criteria. In other words, there was no difference between fair machine guidance and bias feedback in terms of participants' attitudes toward accepting guidance. However, fair machine guidance was more effective at eliciting a change in decision-making criteria.

From the open-ended responses, the following changes in decision-making criteria were identified under fair machine guidance:
\begin{quote}
    \begin{description}
        \item[P48:] Although I did not put much emphasis on ``working hours,'' the advice made me place greater emphasis on it. [Income, Fair machine guidance]
        \item[P64:] Drawing on this advice, I also included race, age, and family structure in the criteria. [Income, Fair machine guidance]
        \item[P72:] The ideas regarding marital status and position in the household were the opposite of mine; however, I was convinced and decided to follow the suggested criteria. [Income, Fair machine guidance]
    \end{description}
\end{quote}

Even in fair machine guidance, 11 of 49 participants did not change their decision-making criteria owing to their distrust of AI or confidence in their answers.
\begin{quote}
    \begin{description}
        \item[P1:] The decision-making criteria did not change. I think the best way to view this is from a human perspective. [Income, Fair machine guidance]
        \item[P45:] I do not think my decisions have changed; rather, the hints convinced me that the type of job, working hours, and other factors I thought important were indeed important. [Income, Fair machine guidance]
    \end{description}
\end{quote}

In the bias feedback group, only one participant changed the criteria independently. The other participants either varied the way they adjusted their numbers for race or gender or did not change them at all:
\begin{quote}
    \begin{description}
        \item[P59:] At first, I focused on the criteria of owing a home and the occupation, but I gradually began to focus on the desired amount (loan) and repayment period. [Credit, Bias feedback]
    \end{description}
\end{quote}
As stated in Section~\ref{subsec:unfairness}, there was no difference in unfairness improvement between fair machine guidance and bias feedback groups, but the change in the decision-making criteria was smaller in the bias feedback group. Many participants in the bias feedback group only controlled how they adjusted their headcounts to improve fairness.

Next, we quantitatively analyzed how the participants used AI-based guidance to change their decision-making criteria. Drawing from the participants' open-ended responses, we investigated how they accepted and utilized the guidance, and classified them into three categories: ``accepted,'' ``rejected,'' and ``confused (wanted to accept but could not utilize the guidance).'' Each of these categories is comprehensively analyzed in the following sections.

\subsubsection{Participants who accepted the guidance} 
The fair machine guidance group included a few participants who simply followed the AI-based guidance and stated that they did this because ``it was the AI's decision'' or because they were unsure of their own decisions. Many participants (21 of 49) accepted the decision-making criteria of the AI system after examining it and being persuaded.
\begin{quote}
    \begin{description}
        \item[P13:] Because the AI-based advice was persuasive, I thought I would follow it. [Income, Fair machine guidance]
        \item[P64:] I referred to the hints provided by the AI because some criteria differed from my own. The criteria provided by the AI were satisfactory; therefore, I did not doubt them. [Income, Fair machine guidance]
    \end{description}
\end{quote}

Additionally, 13 of 49 questioned the AI decision-making criteria but accepted them:
\begin{quote}
    \begin{description}
        \item[P18:] Some of the hints were accurate, whereas some were questionable, and I think I followed some of them in essence and prioritized others. [Income, Fair machine guidance]
        \item[P24:] I was unsure and questioned some of the guidance, such as why ``country of origin'' should be a decision-making criterion, why ``household position'' and ``single/cohabiting'' should be expected to have a higher annual income, or why ``profession'' and ``specialist position'' should indicate lower annual income. Moreover, employment status and working hours are persuasive factors. [Income, Fair machine guidance]
    \end{description}
\end{quote}
The phenomenon of questioning but obeying was unique to the fair machine guidance group. As discussed below, those in the bias feedback group who questioned the guidance did not accept it. Under fair machine guidance, the AI system's presentation of the adjustment method for fair decision-making allowed participants to consider the content and deliberate on the decision-making criteria. As a result, this led to behavior that was not just simplistic acceptance or rejection, but rather a compliance that involved either being convinced or questioning while still following the guidance.

\subsubsection{Participants who rejected the guidance} 
Under fair machine guidance, 11 of 49 participants rejected the AI system's guidance and the most frequently cited reason for this was that the respondent did not agree with the information presented by the AI system after carefully considering its decision-making criteria:
\begin{quote}
    \begin{description}
        \item[P14:] The decision-making criteria of the AI were strange; the left and right panels displayed the same information in different colors, and the system coerced me to use items that were not used in the criteria for the decision; therefore, I did not refer to them much. [Income, Fair machine guidance]
        \item[P90:] It was unclear why the items that I did not focus on were colored, and I questioned whether it would truly be fair if I followed the ``fair criteria.'' There were many unknowns, and I found few parts to be persuasive. [Income, Fair machine guidance]
        
        \item[P7:] I was not sure of what constitutes fairness. I could not understand on what basis it is considered fair. [Credit, Fair machine guidance]
    \end{description}
\end{quote}

Interestingly, even those who indicated that they did not follow the AI guidance adjusted their decision-making criteria based on their awareness of their own criteria.
\begin{quote}
    \begin{description}
        \item[P15:] I was not persuaded by the hints presented (and did not follow them). I felt that (for me) there was a tendency to judge one's ability to pay based on their occupation. I thought I was making a choice while trying not to make decisions based on race or gender, but when I was at a loss. I felt that I would eventually filter my decision based on ``how trustworthy'' how the person seemed based on factors such as their appearance. [Credit, Fair machine guidance]
    \end{description}
\end{quote}
As we have already observed, even if one does not agree with or trust AI, each step in the mental correction process prompts a reconsideration of fairness, awareness of one's own criteria, and adjustment of their decision-making criteria.

Some participants from the bias feedback group questioned the need to adjust the number of people to make fair decisions. Those with such doubts responded that they did not use the AI guidance but instead made their own decisions. This contrasts with the fair machine guidance group, in which some participants followed the AI guidance despite their doubts.
\begin{quote}
    \begin{description}
        \item[P31:] It felt cumbersome to adjust the numbers while thinking about race, rather than purely about my own income expectations, and I was left wondering ``just what is fairness in the first place?'' when matching numbers based on race rather than ability. [Income, Bias feedback]
    \end{description}
\end{quote}

Note that fair machine guidance also provided a bias score, but no questions were raised regarding the numerical value. Participants in the fair machine guidance group were more concerned with the visualization of decision-making criteria than with numerical values.

\subsubsection{Participants who were confused by the guidance}
After receiving fair machine guidance, 4 out of 49 participants were willing to accept the decision-making criteria offered by the AI system but were unable to use them in practice. In particular, some participants did not understand the criteria presented by the AI and its decision-making process. Additionally, many were confused because there was excessive information in the given AI criteria, and they could not comprehend what to prioritize and use to aid their decisions.
\begin{quote}
    \begin{description}
        \item[P75:] There were multiple decision items and I was unsure which one to prioritize. [Income, Fair machine guidance]
        \item[P60:] I wanted to follow the advice but could not because I did not understand why the AI provided such a result. [Income, Fair machine guidance]
    \end{description}
\end{quote}

In the bias feedback group, similar to the fair machine guidance group, some participants stated that they could not use the information provided because they did not understand it. However, unlike in the fair machine guidance group, most participants in the bias feedback group stated that they did not know how to change their decisions to make them fair based only on the numerical information provided regarding unfairness: ``I tried to follow the AI system's advice but not know how to do so'' (P32, [Income, Bias feedback]).

The fact that bias feedback alone was insufficient to encourage a change in the decision-making criteria because the participants were not familiar with how to change their decisions confirms the importance of guidance for decision-making criteria in fair machine guidance. However, fair machine guidance can sometimes cause confusion.

%% file: source/body/discussion.tex
\section{Discussion}
\subsection{Challenges in motivating bias mitigation}
We examined the impact of AI teaching on human decision-making by focusing on the mental correction process.
Our study revealed that fair machine guidance, which prompts users to re-evaluate and adjust their decision-making criteria for fairness, effectively encourages them to reconsider their notions of fairness and motivates them to address their biases. Interestingly, even when users were skeptical about the fairness criteria provided by the AI system, they did not outright reject them. Instead, they interpreted the AI guidance to be reflective of societal expectations of fairness, concluding that adopting diverse perspectives is crucial.

Moreover, fair machine guidance did not prompt the rethinking of fairness in some participants because it involved guiding only fair decision making and did not stress the importance of impartiality. Hence, fair machine guidance can be used in combination with other methods, such as using lectures, to teach the importance and necessity of fairness to instill further motivation~\cite{greenwald1998measuring}. Moreover, some of the open-ended comments by the participants did not indicate the importance or necessity of fairness in their real lives. Some participants stated that they were not familiar with this concept because they lived in an environment wherein racism was not readily perceived. Therefore, experiencing the importance of impartiality through role-playing may be useful for creating opportunities to reconsider impartiality. For example, role-playing methods were applied to motivate participants to act on misinformation and gain experience in dealing with it~\cite{Roozenbeek2019-eg,Chang2020-be}.

\subsection{Challenges in raising awareness of the direction and magnitude of biases}

The experimental results revealed that fair machine guidance fosters introspection into one's own decision-making criteria and enhances awareness of specific biases, such as making decisions based on certain information. Additionally, existing bias-mitigation approaches have exhibited limited effects~\cite{Lichtenstein1980-vi}. It was reported that when bias correction was targeted at a specific bias and feedback specific to that bias was provided, users improved their decision-making, but they applied a simplistic strategy. In our study, this phenomenon was mainly observed under the bias feedback condition. In other words, this method tends to be simplistic, as it focuses on sensitive attributes to prompt fair decisions. In contrast, fair machine guidance promotes deep reflection on the significance of decision making based on diverse perspectives. This approach has the potential to positively impact not only on personal assessments but also various decisions in daily life.

Furthermore, we observed that not all participants completely understood the information presented by the AI system. This may be because of the amount of information presented in the criteria. Under fair machine guidance, we visualized and presented both the participants' own and fair decision criteria. However, the volume of this information confused some participants, as they did not know what to prioritize. We do not consider this confusion negative but rather a step forward in the process of making fair decisions. Additionally, in the bias feedback group, we only presented numerical information regarding biases. However, many participants felt that this was insufficient and did not know how to revise their decisions. Therefore, in the future, the quantity and quality of the information presented to humans should be addressed.

\subsection{Challenges in adjusting decision criteria}
Fair machine guidance encouraged participants to change their decision-making criteria toward achieving fair evaluations. In particular, their decision criteria changed such that they focused on a wider range of information. Although some participants accepted the AI system's guidance and adjusted their decision-making criteria, others did not agree with the fairness criteria presented. There were many cases in which participants deliberated on fair decision-making and amended their own decision-making criteria by comparing their decisions with those presented by the AI system. From these results, we believe that the key to encouraging people to adjust their decision-making criteria is not to convince them to accept the AI guidance, but instead to allow them to go through the process of deliberating the information presented by it.

However, fair machine guidance could not suppress response coordination. Even when it is possible to correct decision-making biases, overcorrection may occur~\cite{Harber1998-fg,Petty1997-so}, which was also observed in this study. This is because although fair machine guidance suggested that the criteria should be revised to make a fair decision, the participants were responsible for identifying the extent to which they should be modified. Most of them prioritized the unprivileged group (i.e., non-whites or females) to mitigate unfairness, which would result in reverse discrimination against the privileged group if performed excessively. For fair decision-making, precautions must be taken against excessive adjustments to prevent reverse discrimination. In previous studies, the excessive modification of decision-making bias, known as the rebound effect, was confirmed in various domains~\cite{Galinsky2007-qe,Monteith1998-qk,Oe2003-wb,Ko2008-gw}. Various efforts, including thought suppression~\cite{Galinsky2007-qe} and replacement thinking~\cite{Hirt1995-tm,Oe2003-wb}, have been made to reduce these effects. The approach of thought suppression avoids unwanted thoughts such as stereotypes. For example, stereotyping can be suppressed by postponing or not thinking about speaking when expressing discriminatory opinions. Alternative thinking suppresses stereotypes by rethinking content differently. For example, to suppress the stereotype ``Black people are aggressive'', alternative thinking is used to form opposing or neutral opinions, such as ``Black people are honest'' ~\cite{Galinsky2007-qe}. Studies showing reduced rebound effects have focused on whether alternative thinking is easy, and it is important to consider the contents of alternative thoughts.
We believe that incorporating these approaches to suppress rebound effects in this study will be useful.

\subsection{Relation to algorithm aversion}

Previous discussions on AI decision support have focused on addressing algorithm aversion, which is more pronounced when the subject of the decision is subjective to the evaluator~\cite{Schecter2023-ju,De-Arteaga2020-in,Wang2020-dk,Dietvorst2015-ob}. This was also observed in this study and was elicited by fair machine guidance in numerous ways, the first of which involved the differences in fairness criteria. The evaluators' notions of fairness criteria and those taught by the AI system contained divergences, thereby creating aversion. This is similar to the phenomenon noted in previous studies on aversion resulting from differences between one's own ideas and those of AI systems~\cite{De-Arteaga2020-in,Schecter2023-ju,Burton2020-qd}. Prior studies have summarized why evaluators are reluctant to be persuaded to correct their biases; for example, one reason for reluctance is that they dislike having their decisions questioned and do not want their decision-making to be manipulated~\cite{Larrick2004-ap}. Brehm et al. found that methods for mitigating bias, especially those based on external feedback, can create a backlash because they are perceived as indirectly denying the evaluators the freedom to decide~\cite{Brehm1966-di}. Similar situations were also identified in our study with open-ended comments such as ``I think AI is incapable of making fair decisions,'' and ``I feel uneasy about being strongly armed by AI.'' The results of our experiment suggest that in AI-based education, deliberation on the information presented by AI is more important than trust in AI. Nonetheless, it is important that evaluators trust and use AI. Although this study specifically used AI as a tool for bias correction, methods and strategies aimed at persuading humans to use AI should be considered when applying it to other subjects and domains.

The second reason relates to the information presented. Previous AI-based decision-making support has adopted the approach of gaining human trust by presenting the decision-making criteria of the AI system~\cite{poursabzi2021manipulating,hase2020evaluating,nguyen2018comparing,Wang2021-lk,Ma2023-bi,Wang2021-yd,Park2022-sd}. Based on this idea, we attempted to visualize the decision-making criteria of our AI system to help participants understand them. However, this visualization resulted in algorithm aversion among some participants, with responses indicating that the participants attributed their distrust to the fact that the information to which they had not paid attention was emphasized. Additionally, the emphasis on sensitive information, particularly  gender and race, was disliked by many participants. However, the clarification of the decision-making criteria had a positive impact on some participants. For example, it encouraged them to consider and heed new perspectives. Thus, while visualizing the decision-making criteria induced algorithm aversion in decision making, the benefits of this approach were also indicated.

\subsection{Application to other domains}\label{subsec:appfordomain}
The AI guidance to improve decision making explored in this study for fair person assessment could also be beneficial in other domains, such as those having societal consensus on the appropriate fairness standard. For instance, as mentioned in Section~\ref{subsec:fairness}, employment and loan evaluations are legally established fairness standards in some countries. Additionally, the medical field has clear goals and guidelines for adolescent health, as outlined in~\cite{Richmond2006-aq}.

In contrast, domains where AI guidance for decision-making is not advisable include those with unclear fairness standard. For instance, a participant expressed concerns about potential social confusion due to a lack of consensus on what constitutes a correct decision: ``No one truly knows if it (the fair AI) is truly correct. Despite this, many people may blindly believe in it, potentially causing social confusion'' (P45, [Income, fair machine guidance]). Examples of such domains include prioritizing organ transplants in medical settings or legal interpretations. In these domains, the use of machine guidance with the assumption that it provides the absolute correct answer is not recommended. Once consensus on the fairness standard is established, fair machine guidance has the potential to support fair decision-making in such domains. For example, jury trials, which deliberately seek judicial decisions from individuals without specialized training, may increasingly require fair judgments in the future. Fair machine guidance could be a tool for guiding individuals in such situations.

\subsection{Expectations from and concerns regarding fair machine guidance}

Finally, based on the open-ended questions, we discussed the type of AI required in current society. First, we summarized the need for AI assistance (responses to Q13): Many participants felt that they required assistance but did not want to entrust AI with the entire decision-making process. We found that many participants sought AI support when they were unsure or unaware of their decisions. Furthermore, existing research shown that decision support requires AI to make rational decisions based on large amounts of data~\cite{Kapania2022-gx,Lima2021-cc,Sundar2019-cj,Park2022-sd}. By contrast, those who felt that they did not require AI assistance emphasized human touch and expressed dissatisfaction with the accuracy of current AI technology as the main reason. In particular, opinions that emphasize the role of human decision-making in fairness suggest that human nature and initiatives will retain their value regardless of the development of AI technology.

We then summarized the concerns regarding AI support for practical use (responses to Q14). The participants raised numerous concerns regarding the practical application of the AI system. The first was the limitations of AI capabilities. Some participants believed that regardless of how technology evolved, AI would not be able to reproduce human sensibilities. Additionally, they did not believe that even AI could be perfectly impartial and were concerned that it would make incorrect decisions. The next concern was over-reliance on AI, as many participants were concerned that it would reduce human decision-making ability. There were also concerns that heavy reliance on AI may lead to the loss of human ethics and sensibilities. Finally, we reviewed concerns regarding the misuse of AI. Many participants raised concerns regarding the misuse of AI by malicious actors. In particular, the AI users were concerned that it was difficult to identify signs of misuse and that they might be unknowingly guided.

These comments indicate that people do not want AI to be a complete decision-making replacement system but rather as a tool to support their own decision-making. These findings support the idea that, even with AI for decision-making support, final decision-making should be the prerogative of humans~\cite{Panigutti2022-xu,Chiang2023-tm}. Additionally, as popularity of AI increased, some participants have become concerned about the deterioration of decision-making because of over-reliance. This concern suggests that our approach to AI-based education is necessary, as it is not only involves providing information and advice but it aims to improve the decision-making ability of an individual. 

Concerns regarding the misuse of AI have also been noted. Boosting, an intervention to enhance cognitive processes, is known to be effective as an approach to counteract the manipulative influence of AI. For example, boosting, which raises awareness of one's own personality, can increase the awareness of the misuse of micro-targeting advertising~\cite{Lorenz-Spreen2021-wu}. Reflective writing is one form of boosting and involves individuals writing out their judgments and reasoning~\cite{Krsek2022-gl}. This approach encourages consideration and self-reflection and aligns with the insights gained from our results. Integrating reflective writing into our system can help users critically assess and appropriately utilize AI guidance instead of accepting it blindly. AI system developers should also enhance the transparency and interpretability of their systems to prevent inadvertent harm to users.

%% file: source/body/conclusion.tex
\section{Limitations and future work}

\subsection{Choice of fairness metrics}
We used demographic parity as the fairness metric because it is commonly used in the dataset employed for the experiment and is a fairness standard mandated by law in some countries. We believe that indicators for guiding humans, including fairness metrics, should be  provided externally based on societal consensus~\cite{Cheng2021-gn}. In addition to demographic parity adopted in this study, there are different types and purposes of fairness metrics, such as equalized odds~\cite{hardt2016equality} and individual fairness~\cite{Dwork2012-va}. The findings and conclusions of this study cannot be simply generalized to other fairness metrics or contexts. 
For instance, it is known that equalized odds is generally more difficult to understand than demographic parity. This is mainly because concepts used in equalized odds, such as the false negative rate and the false positive rate, are less familiar to the citizens~\cite{Saha2020-xx}. When there is a lack of understanding about such fairness metrics, it will likely become more challenging for people to reassess their views on fairness and modify their decision-making criteria in the same manner as with demographic parity.

Although fairness metrics are socially and legally defined in areas such as employment and loan evaluations, there are cases in which no societal consensus exists, and individuals hold different beliefs regarding what constitutes fairness. Indeed, some participants felt that the fairness metric we set was differed from their perceptions of fairness. We see potential in the AI guidance to re-evaluate and form a consensus on fairness metrics in cases where such an agreement does not exist. This potential is supported by the results that show our system can encourage people to reconsider the fairness metric. Although our experiment utilized a single fairness metric, involving participants in rethinking fairness using multiple metrics and then aggregating their opinions or facilitating discussions could contribute to the formation of a consensus. However, a lack of understanding of fairness metrics could hinder this re-evaluation, indicating the need for approaches that enhance the understanding of each fairness metric.

\subsection{Collective decision-making}
In this study, we focused on individual evaluations. However, in real-world scenarios, evaluations are often performed collectively and diversity is crucial for improving group decision-making~\cite{Fujisaki_Honda_Ueda_2018,Kurvers2016-tp,Novaes_Tump2018-as}. The proposed method, which creates a teacher model based on each user's judgment, maintains diversity. However, during consensus building through discussions, a wide range of diverse judgments may be overlooked, resulting in decisions based solely on a few individuals' opinions. In our experiments, the extent of improvement in fairness varied among the participants. When the opinions of those with less impact on improving fairness are overly represented, the collective decision-making can become biased. To ensure fair group decision-making, combinations of methods that aggregate opinions while considering diverse values, such as those proposed in~\cite{Baba2020-qq}, should be employed.

\section{Conclusion}
In this study, we explored the impact of fair machine guidance on human decision-making. In fair machine guidance, a fair model trained through fairness-aware ML acts as a teacher to guide humans in fair decision-making. Although the experimental results did not indicate a significant reduction in unfairness when using fair machine guidance compared to the simple method, they revealed notable differences in the participants' process of improving fairness. Specifically, fair machine guidance prompted them to rethink their fairness through motivation for bias correction. Moreover, it also increased their awareness of their own biases and the importance of diverse perspectives. We confirmed that this approach was effective at encouraging participants to adjust their decision-making criteria.

Unlike in previous AI-based decision assistance, our results indicated that the participants neither blindly accepted nor dismissed the AI input; instead, they carefully evaluated the information provided. These results suggest that meaningful reflection on AI-provided information is more crucial for informed decision making than simply relying on it. 
Based on our findings and those of previous studies, we argue that presenting information about the decision-making process, not just the final answers, enables individuals to make more thoughtful decisions.

%% file: source/main.bbl

%% file: source/body/appendix.tex
\onecolumn

\section{Appendix}
\subsection{Questionnaires}

\begin{table}[h]
\caption{Questionnaire after the pre-test}
\label{tbl:pre-questionnaire}
\begin{tabular}{|p{14cm}|}
\hline
Q1. Do you think that the percentage of people who receive favorable judgments in rental checks and credit checks by financial institutions (e.g., people who are allowed to move in or borrow money) should be the same regardless of race? [5-point Likert scale]\\
\hline
Please elaborate on the reasons for your answer to Q1. [text]\\
\hline
Q2. Do you think your decisions were fair in the 10 questions you just answered? [5-point Likert scale]\\
\hline
Please elaborate on the reasons for your answer to Q2. [text]\\
\hline
Q3. In the 10 tasks you just completed, please tell us up to five attributes that you valued while predicting annual income. [multiple choice] \\
\hline
Please tell us why you valued the attributes selected in Q3. [text] \\
\hline
\end{tabular}
\end{table}

\begin{table}[h]
\caption{Questionnaire after the post-test; All participants in the fair machine guidance (FMG) condition answered all the questions, while those in the bias feedback condition (BF) only responded to some of them.}
\label{tbl:post-questionnaire}
\begin{tabular}{|c|c|p{13cm}|}
\hline
  BF                 & FMG                  &  Questionnaire \\  \hline
 \multirow{1}{*}{\checkmark}    & \multirow{1}{*}{\checkmark}            &  Q4. Do you think your decisions were fair in the last 10 questions you answered?  {[}5-point Likert scale{]}                                                                                                   \\ \hline 
 \multirow{1}{*}{\checkmark}    & \multirow{1}{*}{\checkmark}            &  Please elaborate on the reasons for your answer to Q4. {[}text{]}                                                                                                                                            \\ \hline 
 \multirow{2}{*}{\checkmark}    & \multirow{2}{*}{\checkmark}            &  Q5. We presented the machine-generated advice to help you be fair in your judgment. Did you decide to follow the advice? {[}5-point Likert scale{]}                                                                      \\ \hline 
 \multirow{1}{*}{\checkmark}    & \multirow{1}{*}{\checkmark}            &  Please elaborate on the reasons for your answer to Q5. {[}text{]}                                                                                                                                            \\ \hline 
                                & \multirow{2}{*}{\checkmark}            &  Q6. We presented the judgment example of a ``Fair AI'' similar to the one shown below to help you be fair in your judgment. Did you understand why ``Fair AI'' made such judgments? {[}5-point Likert scale{]} \\ \hline 
                                & \multirow{1}{*}{\checkmark}            &  Please elaborate on the reasons for your answer to Q6. {[}text{]}                                                                                                                                            \\ \hline 
                                & \multirow{1}{*}{\checkmark}            &  Q7. Do you think that the ``Fair AI'' was fair? {[}5-point Likert scale with the ``I don't know'' option{]}                                                                                                    \\ \hline 
                                & \multirow{1}{*}{\checkmark}            &  Please elaborate on the reasons for your answer to Q7. {[}text{]}                                                                                                                                            \\ \hline 
                                & \multirow{2}{*}{\checkmark}            &  Q8. Do you think that ``your criteria'', which were presented in the guidance matched your actual criteria? {[}5-point Likert scale{]}                                                                           \\ \hline 
                                & \multirow{1}{*}{\checkmark}            &  Please elaborate on the reasons for your answer to Q8. {[}text{]}                                                                                                                                            \\ \hline 
 \multirow{2}{*}{\checkmark}    & \multirow{2}{*}{\checkmark}            &  Q9. Did you have any doubts about the advice presented? Alternatively, were there any with which you agreed? Please specify. {[}text{]}                                                                 \\ \hline 
 \multirow{2}{*}{\checkmark}    & \multirow{2}{*}{\checkmark}            &  Q10. Did these tasks cause you to reconsider the fairness of your own decision and that required by society? {[}5-point Likert scale{]}                                                                \\ \hline 
 \multirow{2}{*}{\checkmark}    & \multirow{2}{*}{\checkmark}            &  Please elaborate on the reasons for your answer to Q10. In particular, if you answered 4 or 5, then please elaborate upon your thoughts about fairness. {[}text{]}                                                \\ \hline 
 \multirow{2}{*}{\checkmark}    & \multirow{2}{*}{\checkmark}            &  Q11. In the 10 tasks you just completed, please tell us up to five attributes that you valued when predicting annual income. {[}multiple choice{]}                                                      \\ \hline 
 \multirow{1}{*}{\checkmark}    & \multirow{1}{*}{\checkmark}            &  Please tell us why you valued the attributes selected in Q11.                                                                                                                                           \\ \hline 
 \multirow{1}{*}{\checkmark}    & \multirow{1}{*}{\checkmark}            &  Q12. Did your judgment criteria change after receiving the advice? Please specify. {[}text{]}                                                                                                           \\ \hline 
 \multirow{2}{*}{\checkmark}    & \multirow{2}{*}{\checkmark}            &  Q13. Do you think AI is necessary to help you make fair judgments? If yes, please specify the situations in which you would like to use it. {[}text{]}                                                  \\ \hline 
 \multirow{2}{*}{\checkmark}    & \multirow{2}{*}{\checkmark}            &  Q14. Would you feel concerned if the AI that helps you make fair judgments becomes available for practical use? If you do have concerns, please specify them. {[}text{]}                                \\ \hline 
 \multirow{1}{*}{\checkmark}    & \multirow{1}{*}{\checkmark}            &  Please state your race? {[}Asian/White/Black or African American/Prefer not to say{]}                                                                                                                 \\ \hline 
 \multirow{1}{*}{\checkmark}    & \multirow{1}{*}{\checkmark}            &  Please state your gender? {[}Female/Male/Non-binary/Prefer not to say{]}                                                                                                                              \\ \hline 
 \multirow{1}{*}{\checkmark}    & \multirow{1}{*}{\checkmark}            &  Please state your age? {[}--19/20--29/30--39/40--49/50--59/>60/Prefer not to say{]}                                                                                                                        \\ \hline

\end{tabular}
\end{table}